\RequirePackage{fix-cm}
\documentclass[twocolumn,epjc3]{svjour3_modified}  
\smartqed  % flush right qed marks, e.g. at end of proof
\RequirePackage{graphicx}
\RequirePackage[colorlinks,citecolor=blue,urlcolor=blue,linkcolor=blue]{hyperref}

\usepackage[acronym]{glossaries}
\usepackage{amsmath,amssymb,amsfonts}%
\usepackage{mathrsfs}%
\usepackage[title]{appendix}%
\usepackage{xcolor}%
\usepackage{textcomp}%
\usepackage{manyfoot}%
\usepackage{booktabs}%
\usepackage{algorithm}%
\usepackage{algorithmicx}%
\usepackage{algpseudocode}%
\usepackage{listings}%
\usepackage{subcaption}
\usepackage{lmodern}
\usepackage{lineno}
\journalname{Eur. Phys. J. C}
\begin{document}

\title{Long term study of sedimentation and biofouling at Cascadia Basin, the site of the Pacific Ocean Neutrino Experiment%\thanksref{t1}
}

\author{
O.~Aghaei\textsuperscript{1} \and
M.~Agostini\textsuperscript{2} \and
S.~Agreda\textsuperscript{1} \and
A.~Alexander Wight\textsuperscript{2} \and
P.~S.~Barbeau\textsuperscript{3,4} \and
A.~J.~Baron\textsuperscript{1} \and
S.~Bash\textsuperscript{5} \and
C.~Bellenghi\textsuperscript{5} \and
B.~Biffard\textsuperscript{1} \and
M.~Boehmer\textsuperscript{5} \and
M.~Brandenburg\textsuperscript{5} \and
D.~Brussow\textsuperscript{1} \and
N.~Cedarblade-Jones\textsuperscript{3,4} \and
M.~Charlton\textsuperscript{1} \and
B.~Crudele\textsuperscript{2} \and
M.~Danninger\textsuperscript{6} \and
F.~C.~De~Leo\textsuperscript{1,7} \and
T.~DeYoung\textsuperscript{8} \and
F.~Fuchs\textsuperscript{9} \and
A.~G\"{a}rtner\textsuperscript{6} \and
J.~Garriz\textsuperscript{8} \and
D.~Ghuman\textsuperscript{6} \and
L.~Ginzkey\textsuperscript{5} \and
V.~Gousy-Leblanc\textsuperscript{5} \and
D.~Grant\textsuperscript{6} \and
A.~Grimes\textsuperscript{6} \and
C.~Haack\textsuperscript{10} \and
R.~Halliday\textsuperscript{9} \and
D.~Hembroff\textsuperscript{1} \and
F.~Henningsen\textsuperscript{6} \and
J.~Hutchinson\textsuperscript{1} \and
R.~Jenkyns\textsuperscript{1} \and
S.~Karanth\textsuperscript{11} \and
T.~Kerscher\textsuperscript{5} \and
S.~Kershtein\textsuperscript{1} \and
K.~Kopa\'{n}ski\textsuperscript{11} \and
C.~Kopper\textsuperscript{10} \and
P.~Krause\textsuperscript{6} \and
C.~B.~Krauss\textsuperscript{12} \and
I.~Kulin\textsuperscript{1} \and
N.~Kurahashi\textsuperscript{13} \and
C.~Lagunas Gualda\textsuperscript{5} \and
A.~Lam\textsuperscript{1} \and
T.~Lavallee\textsuperscript{1} \and
K.~Leism\"{u}ller\textsuperscript{5} \and
R.~Li\textsuperscript{5} \and
S.~Loipolder\textsuperscript{5} \and
A.~Maga\~{n}a Ponce\textsuperscript{9} \and
S.~Magel\textsuperscript{5} \and
P.~Malecki\textsuperscript{11} \and
G.~G.~Marshall\textsuperscript{2} \and
T.~Martin\textsuperscript{12} \and
S.~Mihaly\textsuperscript{1} \and
C.~Miller\textsuperscript{6,14} \and
N.~Molberg\textsuperscript{12} \and
R.~Moore\textsuperscript{12} \and
B.~N\"{u}hrenb\"{o}rger\textsuperscript{5} \and
B.~Nichol\textsuperscript{6} \and
W.~Noga\textsuperscript{11} \and
R.~\O{}rs\o{}e\textsuperscript{5} \and
L.~Papp\textsuperscript{5} \and
V.~Parrish\textsuperscript{8} \and
M.~Paulson\textsuperscript{1} \and
P.~Pfahler\textsuperscript{5} \and
B.~Pirenne\textsuperscript{1} \and
E.~Price\textsuperscript{1} \and
A.~Rahlin\textsuperscript{15,16} \and
M.~Rangen\textsuperscript{12} \and
E.~Resconi\textsuperscript{5} \and
C.~Ridsdale\textsuperscript{1} \and
S.~Robertson\textsuperscript{12} \and
A.~Round\textsuperscript{1} \and
D.~Salazar-Gallegos\textsuperscript{8} \and
A.~Scholz\textsuperscript{5} \and
L.~Schumacher\textsuperscript{10} \and
S.~Sharma\textsuperscript{11} \and
C.~Spannfellner\textsuperscript{5} \and
J.~Stacho\textsuperscript{6} \and
I.~Taboada\textsuperscript{17} \and
A.~R.~Thurber\textsuperscript{18} \and
M.~Tradewell\textsuperscript{1} \and
J.~P.~Twagirayezu\textsuperscript{8} \and
M.~Un Nisa\textsuperscript{8} \and
B.~Veenstra\textsuperscript{12} \and
S.~Wagner\textsuperscript{1} \and
C.~Weaver\textsuperscript{8} \and
N.~Whitehorn\textsuperscript{8} \and
L.~Winter\textsuperscript{5} \and
M.~Wolf\textsuperscript{1} \and
R.~Wro\'{n}ski\textsuperscript{11} \and
J.~H.~Wynne\textsuperscript{18} \and
J.~P.~Ya\~{n}ez\textsuperscript{12} \and
A.~Zaalishvili\textsuperscript{3,4} \and
}
\thankstext{e1}{e-mail: btveenst@ualberta.ca}


\institute{{Ocean Networks Canada, University of Victoria, Victoria, BC, Canada} \and
{Department of Physics and Astronomy, University College London, Gower Street, London, WC1E 6BT ,UK} \and
{Department of Physics, Duke University, Durham, NC, 27708, USA} \and
{Triangle Universities Nuclear Laboratory, Durham, NC, 27708, USA} \and
{Physik-department, Technische Universit\"{a}t M\"{u}nchen, D-85748 Garching, Germany} \and {Department of Physics, Simon Fraser University, 8888 University Drive Burnaby, B.C. Canada,V5A 1S6} \and
{Department of Biology, University of Victoria, Victoria BC V8N 1V8, Canada}
\and {Department of Physics and Astronomy, Michigan State University, East Lansing, MI 48824, USA} \and
{Department of Physics, Elmhurst University, 190 S. Propsect Ave, Elmhurst, IL, 60126, USA} \and
{Erlangen Centre for Astroparticle Physics, Friedrich-Alexander-Universit{\"a}t Erlangen-N\"{u}rnberg, D-91058 Erlangen, Germany} \and
{Institute of Nuclear Physics, Polish Academy of Sciences, Kraków, Poland} \and
{Department of Physics, University of Alberta, Edmonton, Alberta, Canada, T6G 2E1} \and
{Department of Physics, Drexel University, 3141 Chestnut Street, Philadelphia, PA, 19104, USA} \and
{Department of Physics and Astronomy, University of Victoria, 3800 Finnery Road, Victoria, BC, Canada, V8P 5C2} \and
{Department of Astronomy and Astrophysics, University of Chicago, 5640 South Ellis Avenue, Chicago, IL, 60637, USA} \and
{Kavli Institute for Cosmological Physics, University of Chicago, 5640 South Ellis Avenue, Chicago, IL, 60637, USA} \and
{School of Physics and Center for Relativistic Astrophysics, Georgia Institute of Technology, Atlanta, GA 30332, USA} \and
{Department of Ecology, Evolution, and Marine Biology, University of California, Santa Barbara, CA, USA}
}

\date{Received: date / Accepted: date}

\newacronym{pone}{P-ONE}{Pacific Ocean Neutrino Experiment}
\newacronym{straw}{STRAW-a}{STRings for Absorption length in Water}
\newacronym{strawb}{STRAW-b}{Strings for Abosrption length in Water b}
\newacronym{onc}{ONC}{Ocean Networks Canada}
\newacronym{tdc}{TDC}{time-to-digital converter}
\newacronym{daq}{DAQ}{data-acquisition}
\newacronym{poc}{POCAM}{Precision Optical CAlibration Module}
\newacronym{pocs}{POCAMs}{Precision Optical CAlibration Modules}
\newacronym{sdom}{sDOM}{STRAW Digital Optical Module}
\newacronym{sdoms}{sDOM}{STRAW Digital Optical Module}
\newacronym{pmt}{PMT}{photomultiplier tube}
\newacronym{pmts}{PMTs}{photomultiplier tubes}
\newacronym{led}{LED}{light-emitting diodes}
\newacronym{rov}{ROV}{remotely operated vehicle}

\maketitle

\begin{abstract}
    STRings for Absorption Length in Water (STRAW)-a and b were pathfinder instruments deployed to characterize the anticipated site of the \acrfull*{pone}, which is a future neutrino telescope that will be located in the North Pacific Ocean. Measurements of the evolution of the optical transmission efficiency from \acrshort*{straw} showed a decline over the detector's lifetime for the upward-facing modules. Video footage of the pathfinders strongly suggested this decline was caused by biofouling and sedimentation.  We measure the effect of biofouling and sedimentation to be a decrease in the transparency of upward-facing optical surfaces over 5 years of operations. A majority of downward-facing optical surfaces, which will dominate \acrshort*{pone}'s sensitivity to astrophysical sources, showed no visible biofouling. Extrapolations motivated by biological growth models estimated that these losses started around 2.5 years after deployment, and suggest a reduction in transparency ranging from 35$\%$ of the original to complete obscuration for the upward-facing modules. Samples of biofouling were taken in order to identify the microbial diversity of these organisms and inform potential intervention strategies. Results of the microbial samples and a candidate anti-biofouling strategy that will be tested on upcoming \acrshort*{pone} instruments are discussed. 
    
    %We present a study of the effects of biofouling and sedimentation on pathfinder instrumentation for the \acrfull*{pone}, which will be a neutrino telescope located in the North Pacific Ocean. \acrshort*{pone} will look for high-energy neutrinos by observing the light produced when these neutrinos interact in the water, detecting and digitizing single photon signals in the ultraviolet-visible range. We measure that biofouling and sedimentation caused a decrease in the transparency of upward-facing optical surfaces over 5 years of operations. A majority of downward-facing optical surfaces, which will dominate 
    %\acrshort*{pone}'s sensitivity to astrophysical sources, showed no visible biofouling. Extrapolations motivated by biological growth models estimated that these losses started around 2.5 years after deployment, and suggest a final equilibrium transparency ranging between 0$\%$ and 35$\%$ of the original for the upward-facing modules. 

    %Measurements of the evolution of the transmission efficiency from \acrshort*{straw} show a decline over the detector's lifetime for the upward-facing modules \cite{Hatch:2023gl}. In this work, the evolution of the transmission efficiency for selected modules are presented using the full 53 month data set. A discussion on insights from the recovery and possible mitigation strategies is then presented. 

\end{abstract}

\section{Introduction} \label{intro}
The \acrfull*{pone} will be a cubic kilometre scale neutrino telescope which will be deployed at a depth of 2660 m in the Cascadia Basin in the North Pacific Ocean \cite{Agostini2020}. The \acrshort*{pone} scientific collaboration is partnered with \acrfull*{onc} and will use their NEPTUNE infrastructure, which consists of over 800 km of fiber-optic and power cables connecting undersea research nodes \cite{ONCNeptune}. The \acrshort*{pone} instrumentation will be connected to a node in the Cascadia Basin. This region is located in a flat segment of an abyssal plain on the Juan de Fuca plate, around 200 km off the coast of Vancouver Island in Canada. The region is not active geologically, with typical current speeds on the order of 3 cm/s, and fluctuations up to 10 cm/s, using data from \acrshort*{onc}'s Oceans 3.0 platform \cite{Oceans3}.

Neutrino telescopes are used to study the neutrino sky \cite{IceCubeScience2016,IceCubeGC2023,AntaresDetector,KM3NeT-Status,BaikalGVDDetector}. This is accomplished by observing the light produced by secondary charged particles when neutrinos interact in the water \cite{Halzen2006}. An array of optical instruments equipped with \acrfull*{pmts}, called optical modules, are used to instrument a volume of water and coincident observation of photons among multiple \acrshort*{pmts} enables neutrino detection. Energy deposited in the detector is estimated based on the total amount of light measured and neutrino flavour identification can be approached based on the spatial characteristics of the observed light. Detector operations are therefore tied to the light sensing efficiency of photo-detectors that make up the optical modules. The full \acrshort*{pone} detector will be made up of 70 one kilometre tall mooring lines, instrumented with optical modules \cite{Agostini2020}. The final geometry and construction schedule for \acrshort*{pone} are yet to be finalized. 
%The clarity of the water where the detector is to be constructed is therefore important. This is characterized by the scattering and absorption of the water.
%%%

Using this detection principle requires that the attenuation length of the water is long enough that it is optically and economically feasible to build a large detector. In order to measure the optical properties of the water, the \acrshort*{pone} collaboration deployed the \acrfull*{straw} pathfinder in the summer of 2018  \cite{Boehmer2019}. Using the \acrshort*{straw} instrument, the attenuation length was measured to be around 28 m at 450 nm with no significant time variation, indicating that the Cascadia Basin was a suitable site for a full scale detector \cite{Bailly2021}. The background rate distribution of ambient light due to bioluminescence was measured and monitored by \acrshort*{straw} since March 2019, with rates varying between 10 kHz and 10 MHz \cite{Bailly2021,KilianThesis}. A second pathfinder, \acrshort*{strawb}, was deployed in 2020 \cite{Rea:2021o3,Holzapfel2024}. Both pathfinder instruments were successfully recovered in the summer of 2023. Assets deployed to the proposed P-ONE location in Cascadia Basin for the purpose of characterizing the site are summarized in table \ref{tab:assets}.

\begin{table}[h]
    \centering
    \caption{Instrumentation deployed to the Cascadia Basin to characterize the site for P-ONE.}
    \begin{tabular}{c|c|c|c}
        Asset & Deployed & Retrieved & Length\\
        \hline
       \acrshort*{straw} line 1 & 2018 & 2023 & 150 m\\
       \acrshort*{straw} line 2 & 2018 & 2023 & 150 m\\
       \acrshort*{strawb} & 2020 & 2023 & 450 m
    \end{tabular}
    \label{tab:assets}
\end{table}

\begin{figure}
        \centering
        \includegraphics[scale=0.4]{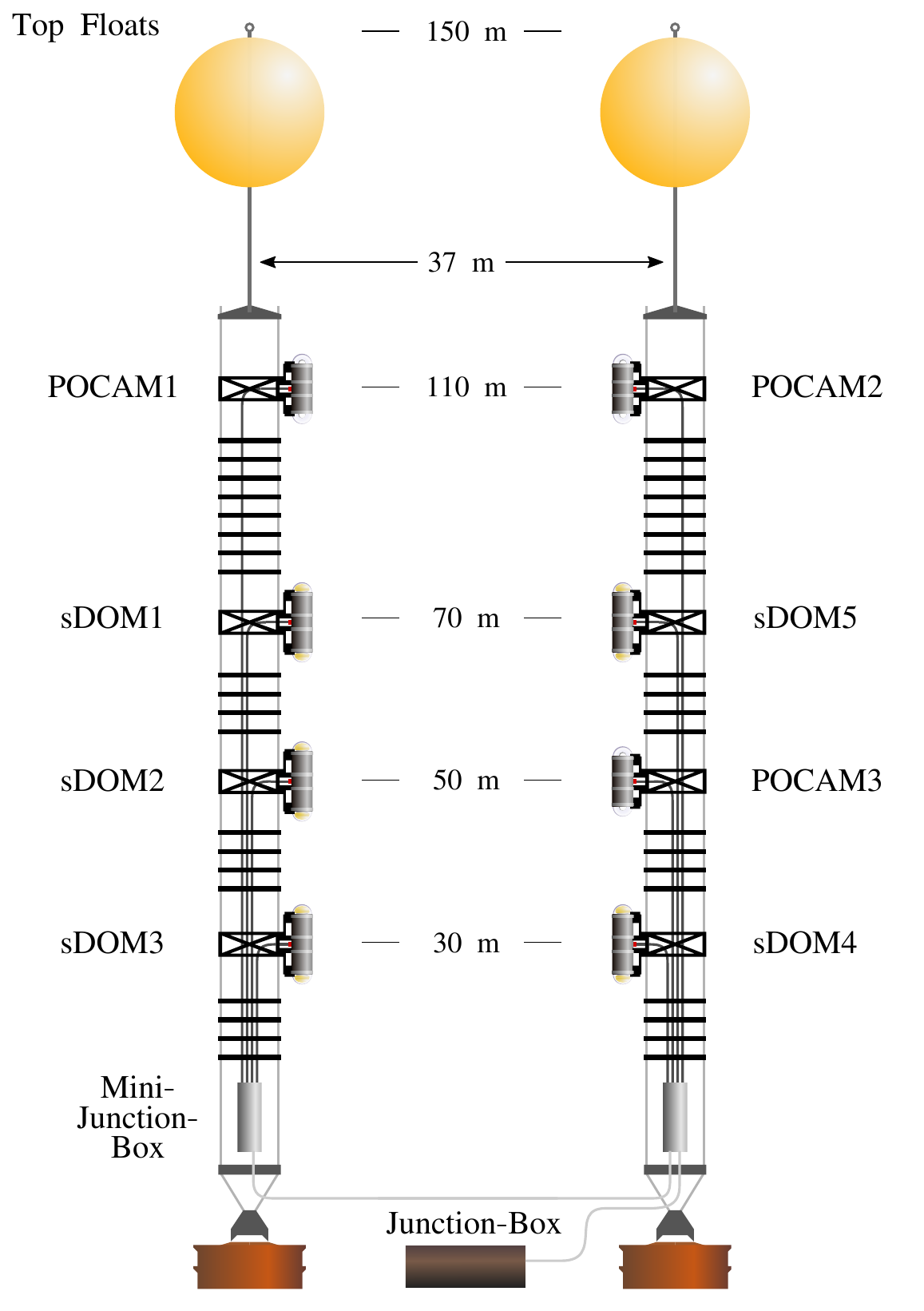}
        \caption{The \acrshort*{straw} instrument \cite{Boehmer2019,Bailly2021}. \acrfull*{poc} modules consisted of different wavelength diodes that could be pulsed at known intensities and acted as flasher beacons \cite{Henningsen2021}. The \acrfull*{sdom}s acted as light collecting modules and consisted of an upward- and downward-facing \acrshort*{pmt}. Figure reproduced from Ref. \cite{Bailly2021} with kind permission of the European Physical Journal (EPJ).} 
        \label{fig:straw}
\end{figure}

Any instrumentation deployed in the ocean environment for long periods of time will be affected by the buildup of marine sediments and biological material, effects collectively known as fouling. 
Marine fouling is caused primarily by three effects and their interactions. Sedimentation refers to inorganic and organic materials vertically sinking from the water column into the seabed \cite{Siegenthaler1993,Hedges1997}. Secondly, a flux of larger particulate aggregates known as marine snow that is composed of detritus, living organisms, organic and inorganic matter, is also transported down from the surface ocean towards the seabed \cite{Fowler1986,Alldredge1988}. Marine snow and sedimentation will be referred to interchangeably, because they both have an equivalent effect of depositing material on submerged surfaces. The third process, known as biological fouling or biofouling, refers to living organisms that colonize submerged surfaces \cite{Gule2016,Carvalho2018,Vuong2023}.

Sedimentation and biological fouling present a concern to neutrino telescopes because the \acrshort*{pmt}s view the water through a pressure resistant glass housing. Material buildup over time makes this glass less transparent, reducing optical module and telescope sensitivity \cite{Salvadori2018}. This effect is strongest on upward-facing surfaces \cite{Amram2003}. 

%The genetic diversity of fouling organisms has previously been found to depend strongly on depth and moderately on substrate orientation \cite{Bellou2012}. The longest such study was done by the ANTARES collaboration, in the Mediterranean Sea \cite{AntaresDetector}. This study found a 20$\%$ average reduction in optical transparency and 15$\%$ detector efficiency reduction over nine years \cite{Salvadori2018}. The ANTARES optical modules faced downwards at a $45^{\circ}$ angle, and were located at depths between 2100 m and 2375 m \cite{AntaresDetector,Amram2002}. ANTARES modules had a reduced exposure to marine snow relative to a surface that was not facing partially downwards.  

\begin{figure}
    \centering
    \includegraphics[scale = 0.3]{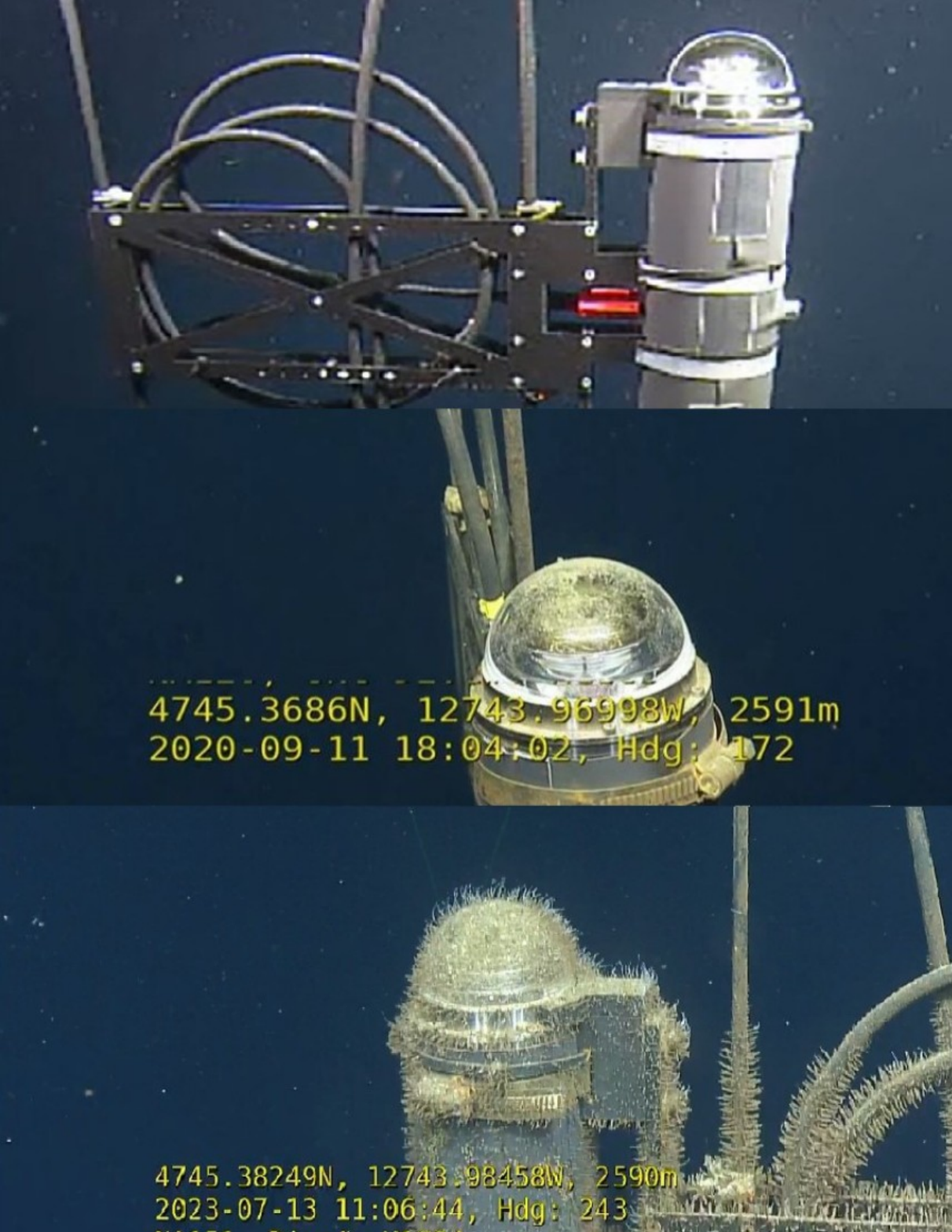}
    \put(-180,230){\color{white}{a.}}
    \put(-180,150){\color{white}{b.}}
    \put(-180,75){\color{white}{c.}}
    \caption{\acrshort*{rov} footage of the upward-facing surface of \acrshort*{sdom}1 on \acrshort*{straw}. The module is shown clean shortly after deployment in 2018 (a.). An amount of buildup is visible in the 2020 survey (b.). The final 2023 survey shows a developed macro-fouling population, composed primarily of hydroids (c.). The downward-facing surface of the same module is shown later, in Fig. \ref{fig: sdom1_recovery}.}
    \label{fig:stacked}
\end{figure}

The \acrshort*{straw} apparatus was visually surveyed three times by means of video from a \acrfull*{rov} deployed from a surface vessel. The first inspection was done in 2018, shortly after deployment. A second survey of \acrshort*{straw} was performed in 2020, showing some material buildup on top of the optical modules \cite{KilianThesis,ImmaThesis}. A third and final survey was done in 2023, just before recovery, that showed an established biofouling population. Figure \ref{fig:stacked} shows the progression of buildup on the upward-facing surface of \acrshort*{sdom}1, located 70 m above the seafloor. Little to no buildup was observed on the majority of downward-facing surfaces, which is discussed in the next section. 

%Measurements of the evolution of the transmission efficiency from \acrshort*{straw} show a decline over the detector's lifetime for the upward-facing modules \cite{Hatch:2023gl}. In this work, the evolution of the transmission efficiency for selected modules are presented using the full 53 month data set. A discussion on insights from the recovery and possible mitigation strategies is then presented.  
The rest of this work concerns analyzing the impact of the observed biofouling buildup on the optical efficiency of the \acrshort*{straw} apparatus. Samples were taken in order to understand the genetic diversity, which could illuminate the mechanism by which biofouling and sedimentation buildup on the instrument and help inform mitigation strategies. \acrshort*{straw} was deployed in late June 2018 and remained on the seafloor for 5 years, however data taking did not start immediately on immersion. Full time data collection started in March 2019, leaving 53 months of data available for the purposes of this study. 

\section{Measuring the Transparency of STRAW Optical Modules}
\label{transparency}

\subsection{The \acrshort*{straw} Detector}
The \acrshort*{straw} instrument consisted of two mooring lines with four instrumented modules each. A schematic of the \acrshort*{straw} apparatus is shown in Fig. \ref{fig:straw}. \acrfull*{pocs} acted as light flashing beacons and contained a precision diffusing sphere with a set of four \acrfull*{led} operating at 365, 405, 465 and 605 nm \cite{Henningsen2021}. The remaining \acrfull*{sdom}s each contained two \acrshort*{pmt}s, one upward-facing towards the surface and the other downward-facing towards the seafloor \cite{Boehmer2019}. 

%Description of STRAW data taking
The \acrfull*{daq} system for each \acrshort*{pmt} on the \acrshort*{sdom} units had two data taking modes. A low-precision mode counted the number of times the \acrshort*{pmt} was triggered in 30 ms intervals and operated continuously \cite{Bailly2021}. The number of \acrshort*{pmt} triggers was converted into a rate. Data from this mode were stored as both a time series, and as histograms of the distribution of rates in one hour intervals, the latter of which were used for this study. 

An alternative high-precision mode could record time over threshold measurements with sub-nanosecond precision using a \acrfull*{tdc} and was run in dedicated campaigns \cite{Boehmer2019,Bailly2021}. Between March 2019 and July 2023, \acrshort*{straw} took data continuously in the low-precision mode with a fractional up-time of 98.3$\%$ in its first two years of operation \cite{Bailly2021}. Across the full 53 month data set, the fractional up-time increased to 99.5$\%$. 

%Description of the data set

%Data taking periods where the \acrshort*{poc} was active were initiated by specifying an intensity (voltage), flasher frequency, and \acrshort*{led} wavelength. The data set used for this study consisted of new measurements taken specifically for the purpose of measuring the efficiency evolution of the optical modules, as well as archival data from previous studies. The majority of archival data had the \acrshort*{poc} set to use the 465 nm LED flashing at a frequency of 2.5 kHz and the intensity set to the maximum of 20 V. New measurements were taken matching these flasher conditions. To compensate for the saturation of the \acrshort*{pmt} nearest the \acrshort*{poc}, the relative efficiencies for the highest \acrshort*{sdoms} were measured using flashes from the \acrshort*{pocs} on the opposing mooring line. 

\subsection{Method Using Low-Precision STRAW Data}
The purpose of the low-precision data taking mode was to measure the rate of background light in the future P-ONE site \cite{Bailly2021}. A broad range of rates extending from a few kHz to rates above the \acrshort*{daq} sampling limit of 10 MHz were observed \cite{Bailly2021}. The lowest observed rates were around 10 kHz and are attributed to light emissions from decays of radioisotopes present in the ocean water, mainly potassium-40, and the intrinsic dark rate of the \acrshort*{pmt}s \cite{Bailly2021}. Very high rates are attributed to the activity of bioluminescent organisms \cite{Bailly2021}. 

\begin{figure}
         \centering
         \includegraphics[scale = 0.45]{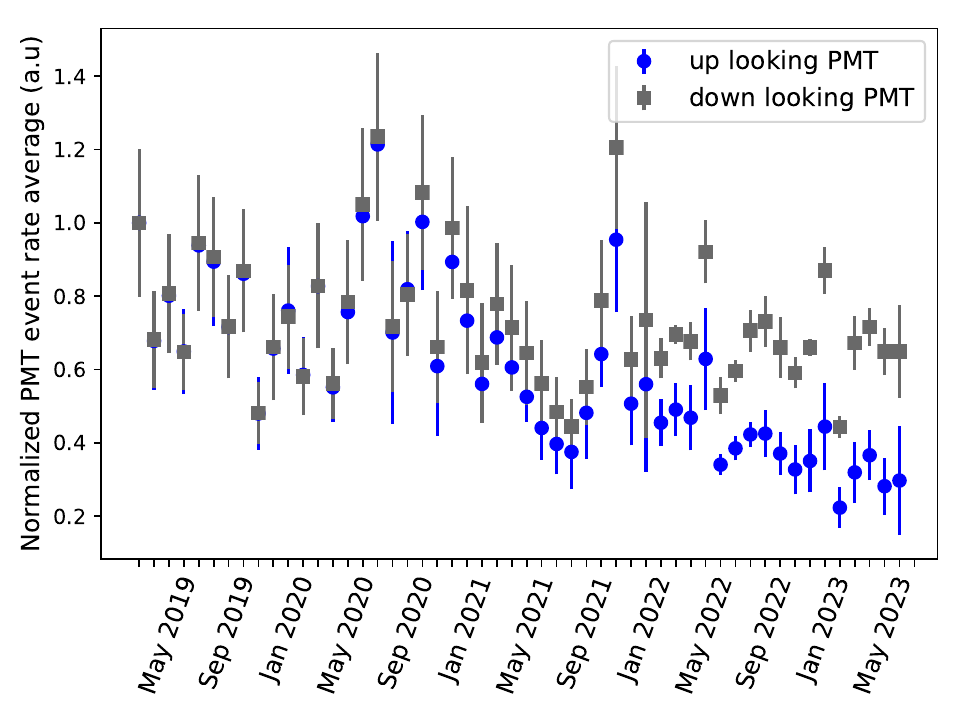}
         \caption{Relative \acrshort*{pmt} trigger rates seen by both upward-facing and downward-facing optical surfaces in \acrshort*{straw} due to ambient light sources, such as bioluminescence. The total light was averaged monthly across all upward- and downward-facing \acrshort*{pmt}s, and then normalized to March 2019. On large time scales, fluctuations in the light measured are observed in correlation between orientations. }
         \label{fig:monthly_correlation}
     \end{figure}

Bioluminescent rates have seasonal variations, which were seen in the \acrshort*{straw} data \cite{Bailly2021}. Figure \ref{fig:monthly_correlation} shows that large monthly fluctuations are consistent between upward-facing and downward-facing \acrshort*{pmt} rates. 

\begin{figure}
       \centering
        \includegraphics[scale = 0.25]{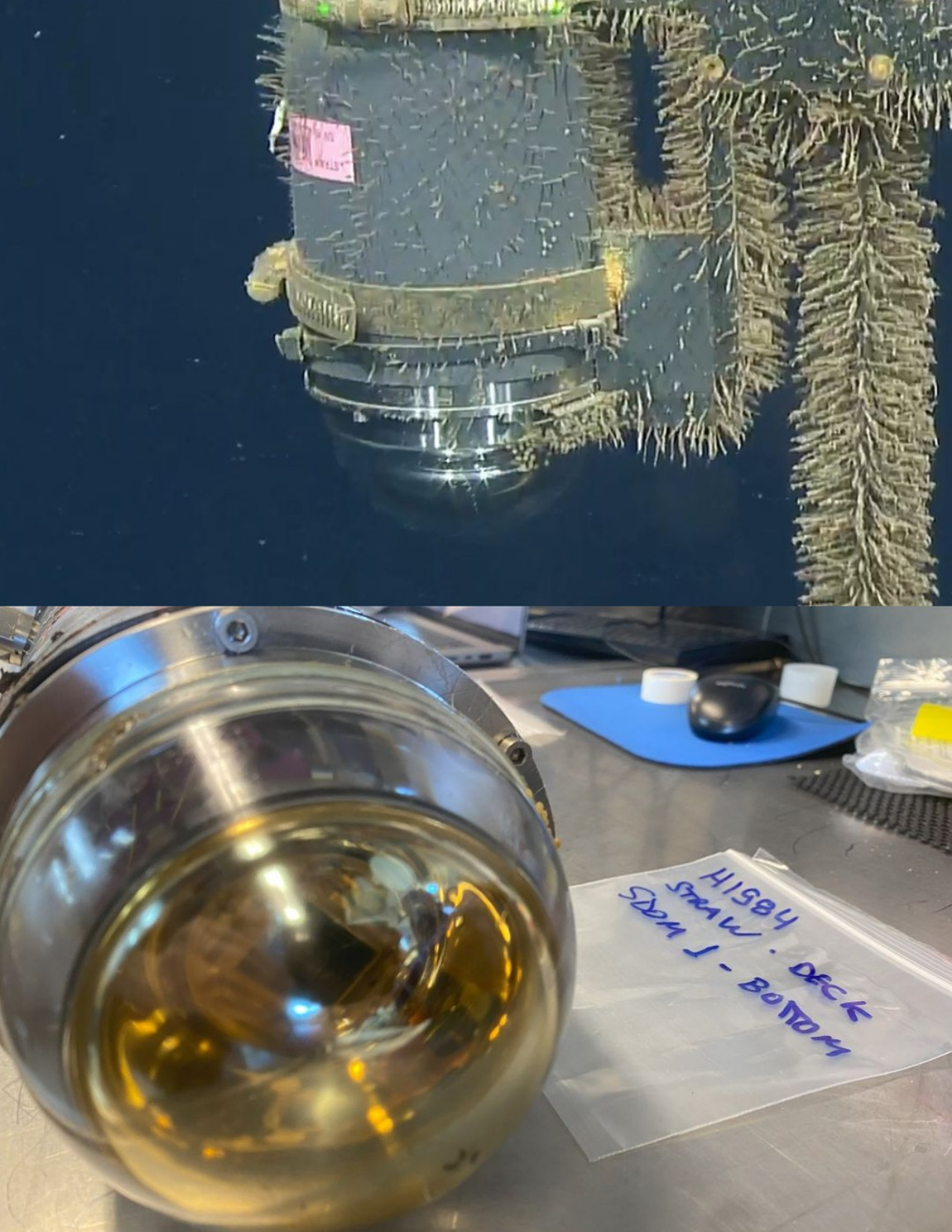}
        \put(-153,192){\color{white}{a.}}
        \put(-153,94){\color{white}{b.}}
        \caption{\acrshort*{straw} \acrshort*{sdom}1 during the \acrshort*{rov} survey (a.) and on deck after recovery (b.), showing an unobstructed downward-facing optical surface.}
        \label{fig: sdom1_recovery}
    \end{figure}

    The majority of downward-facing module surfaces showed little to no visible fouling. Both optical modules positioned 70 m above the seafloor were clear of visible biofouling on the downward-facing glass, one of which can be seen in Fig. \ref{fig: sdom1_recovery}. \acrshort*{sdom}3's downward-facing surface was also observed to be free of visible fouling, at an elevation of 30 m above the sea-floor. The downward-facing flasher modules, located 110 m above the seafloor, were both observed to be free from visible fouling. \acrshort*{sdom}2 and \acrshort*{sdom}4 at 50 m and 30 m above the seafloor did show visible biofouling on the underside. Data from these two modules were excluded from this study, as the method described below could not be used to measure their efficiency over time.  The glass housing \acrshort*{poc}1 and \acrshort*{poc}2, facing down at 110 m above the seafloor, were observed to be free of visible fouling.

   Correlation between rates in up and down looking \acrshort*{pmt}s due to ambient light, combined with the clean downward-facing glass observed in several modules implies that the ratio
   \begin{equation}
       R = \frac{\langle N_{\mathrm{light}} \rangle _{\mathrm{upper}}}{\langle N_{\mathrm{light}}\rangle_{\mathrm{lower}}}, \label{eq:ratio}
   \end{equation}
    is a good estimator for the evolution of the transmission efficiency of the upward-facing optical surfaces. $N_{\mathrm{light}}$ is the average light seen in the binning period, and the average ratio, \textit{R}, was computed monthly for the top most optical modules and one of two modules nearest the seafloor. The result of this \textit{R} measurement is shown in the shaded areas in Fig. \ref{fig:master_plot} where the thickness of the band indicates the $1\sigma$ confidence interval. As a cross check, a selection of fast data was used to independently compute this ratio from bioluminescence and found to be compatible with the results shown in Fig. \ref{fig:master_plot}.

\subsection{High-Precision Method and Results}
Flasher data taking periods used the high-precision \acrshort*{daq} mode. In order to estimate the light collection efficiency, while accounting for attenuation in the water, the fraction of \acrshort*{poc} flashes detected was calculated based on the flasher rate and the measured number of flashes in a one-second integration window. The phase of the time synchronization between the \acrshort*{poc} and the \acrshort*{sdom} was computed using the same method as Ref. \cite{Bailly2021}. This method is not based on both \acrshort*{pmt}s in a module, and therefore wasn't affected by biofouling on the downward-facing \acrshort*{pmt}. 

%consisted of measurements taken specifically for the purpose of measuring the efficiency evolution of the optical modules, as well as archival data from previous studies. The majority of archival data

High-precision data used in this study had the \acrshort*{poc} set to use the 465 nm LED flashing at a frequency of 2.5 kHz and the intensity set to the maximum, as most archival data used in previous studies had used these settings. Efficiencies for \acrshort*{sdoms} 2, 3 and 5 were measured using \acrshort*{poc}1 and \acrshort*{sdom}1 was measured with \acrshort*{poc}2. The reason for using the \acrshort*{poc}s on the opposite mooring line for \acrshort*{sdoms} 1 and 5 was to avoid saturation effects. Communications with \acrshort*{poc}3 were lost in 2020, excluding it from this study.

Only high-precision data where the noise rate was below 100 kHz was used. This cut-off rate was chosen to minimize the chance of a noise hit due to the ambient bioluminescence light occurring in coincidence with the flasher, while retaining a large fraction of the data. In cases where part of an integration window had a background rate above the 100 kHz limit, the high rate portion was excluded and the expected number of flashes was recalculated based on the length of the time window remaining. Data taking periods using the flasher were thirty seconds to one minute long, resulting in 60 nominally one-second integration windows. These 60 measurements were averaged, with the standard error on the mean taken as the statistical error, to compute a data point for each flasher-data-taking period. 

\begin{figure*}
        \centering
        \includegraphics[scale =0.6]{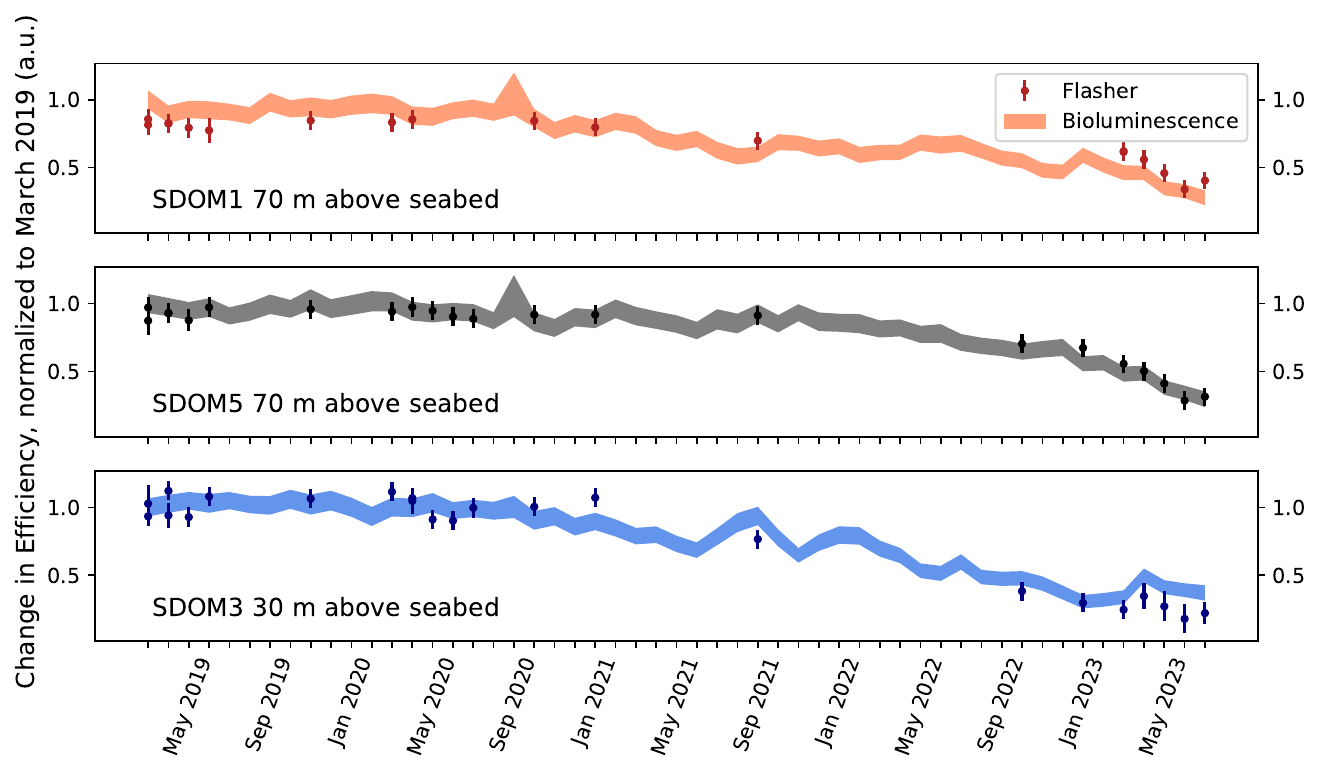}
        \caption{Evolution of the light transmission efficiency of selected upward-facing module surfaces. The shaded region shows the one sigma confidence interval for the ratio obtained using ambient bioluminescence light. Individual points show the results of direct measurements using the flasher. All data is normalized to March 2019, and a scaling factor is used between the two data sets to account for different systematic uncertainties between the two methods. }
        \label{fig:master_plot}
    \end{figure*}

The low-precision data used light generated local to the modules and was averaged monthly, the uncertainties in this data set were assumed to be dominated by statistical fluctuations in the bioluminescence light. The high-precision method could be affected by fluctuations in the attenuation length of the water and distance from the light source, though the measurements presented in Ref. \cite{Bailly2021} do not suggest large time variations. Additionally, individual flasher runs were used to estimate the light collection efficiency, due to data availability, rather than binning over a month. In order to normalize these two different data sets and estimate systematic uncertainties on the high-precision data, a single parameter of the form:
\begin{equation}
    \eta_\mathrm{fast} = s\eta_{\mathrm{slow}}
\end{equation}
was introduced. This single parameter was fit by minimizing a $\chi$-squared function, and the resulting uncertainty was propagated onto the high-precision data points. The results of the different flasher data are shown as points in Fig. \ref{fig:master_plot}.

There was a roughly 8 month period between the deployment of \acrshort*{straw} and commencement of full-time \acrshort*{daq} operation. Some sediment may have built-up on the upward-facing surfaces of the modules during this period. Data points in Fig. \ref{fig:master_plot} that show relative efficiency greater than 1.0 as well as apparent peaks in this data could indicate a cleaning effect due to undersea currents, as was observed in Ref. \cite{Amram2003}. This cleaning effect occurs when above average fluctuations in the undersea current remove loose sediments, restoring the transparency of the module.
    
All data points in Fig. \ref{fig:master_plot} corresponding to measurements using the bioluminescence ratio with efficiency $>$ 1.0 contain 1.0 within their error bars. Two flasher measurements with \acrshort*{sdom}3, corresponding to March 2020 and January 2021 indicate relative efficiency $>$ 1.0. Since this was only observed in one module, we cannot make a definitive statement with regards to this cleaning effect. Apparent peaks in Fig. \ref{fig:master_plot} are most prominent where we do not have flasher data to do high-precision follow up measurements, and therefore any conclusions drawn from these would be speculative in nature. 

The fraction of flashes detected by a \acrshort*{pmt} depends on the attenuation length of the water and distance from the light source. Distance from the light source is affected by ocean currents, and attenuation length may be influenced by seasonal variations in sediment content \cite{Acoustics2025,Seawater1995}. Both are related to tidal effects and therefore expected to be periodic, not continuously decreasing as observed in Fig. \ref{fig:master_plot}. We therefore attribute this loss in transmission efficiency to the accumulation of biofouling and sedimentation that was observed in visual surveys of the \acrshort*{straw} instrument. 

\section{Time Dependence of Fouling Effects}
\label{time_dependence}

The long term behavior is driven by biological effects, which is supported by key observations made during the recovery of the \acrshort*{straw} apparatus. A variety of models and approaches for describing the growth of cells and populations exist in the literature \cite{Charlebois2019}. Most models follow sigmoid shapes and are characterized by three phases: an initial phase with slow growth, an exponential transition and a final state of slow growth where the population asymptotically approaches a limit. A selection of models from the literature were fit to the optical data shown in Fig. \ref{fig:master_plot}. 

A subset of these, the Richards family of models, have been used to analyze a variety of different biological systems \cite{Tjorve2010}. Model selection was assessed on three criteria: goodness of fit, magnitude of uncertainties, and the behavior of models near boundaries. Based on this, two models from the Richards family emerged as candidates, the Logistic and Gompertz models described in Ref. \cite{Tjorve2010} and Ref. \cite{Tjorve2017}. These models are defined by three parameters: the inflection time $T_\textrm{i}$, the relative growth rate at inflection $k_{\textrm{rel}}$, and asymptotic population limit or carrying capacity, $P_{\infty}$. 

Population limit refers to a maximum size or maximum number of the individual units that make up the population. The 2023 \acrshort*{rov} survey identified the larger invertebrates making up the biofouling to be hydroids (Cnidaria: Hydrozoa). Section 4 introduces an analysis of samples taken, revealing the microbial contents of the biofouling. The STRAW apparatus could not be used to count these entities individually. Instead, the assumption was made that the transmission efficiency, $\eta(t),$ would decrease in proportion to the amount of material:
\begin{equation}
    \eta (t)= 1-\alpha P(t), \label{eq: prop}
\end{equation}
where $\alpha$ is a constant of proportionality and $P_\infty$ is the asymptotic population limit. The product of $\alpha$ and $P_{\infty}$ can be fit, we define this as a new parameter
\begin{equation}
    A = \alpha P_{\infty}, \label{eq: alpha}
\end{equation}
representing the fraction of the optical surface that would be obscured by fouling in the long term limit. Applying Eq. \ref{eq: prop} and \ref{eq: alpha} to the functional form of the Logistic model from Ref. \cite{Tjorve2010} gives: 
\begin{equation}
    \eta(t) = 1-\frac{A}{1+e^{-k_{\textrm{rel}}(t-T_\textrm{i})}}.
\end{equation}
Doing the same for the Gompertz model from Ref. \cite{Tjorve2017}:
\begin{equation}
    \eta(t)=1-Ae^{e^{-k_{\textrm{rel}}(t-T_\textrm{i})}}.
\end{equation}

The underlying behavior being modeled is growth on top of a glass substrate. A value of $A$=1.0 implies that the surface becomes opaque in the asymptotic limit. A value of $A>$1.0 would imply that the surface becomes opaque before the growth reaches an equilibrium state. 

The rate, $k_{\textrm{rel}}$ described by this parametrization refers to a relative growth rate \cite{Tjorve2010}. This can be converted to an absolute growth rate, $k_{\textrm{foul}}$, using the equation:
\begin{equation}
    k_{\textrm{foul}} = \frac{\partial P}{\partial t}\big|_{t=T_\textrm{i}}=c_\textrm{m}Ak_{\textrm{rel}}
\end{equation}
where it can be shown that $c_\textrm{m}$ is a model dependent constant for each of the models described in \cite{Tjorve2010}. $c_\textrm{m}$ takes a value of e$^{-1}$ for the Gompertz model and $1/4$ for the Logistic model. Absolute
growth rate may be understood as the maximum rate that the population grows at, or in this case the maximum rate of loss of transmission efficiency. In this work, maximum growth rate is referred to as the fouling rate and is reported for both models in table \ref{tab:mc_bio}.

Fitting was accomplished by minimizing a negative log-likelihood function, assuming Gaussian errors on the \acrshort*{straw} measurements, parameterized as:
\begin{equation}
    -\log L =\sum_{i=1}^n\log \Big(\frac{\eta_{\mathrm{obs}}(t_i)-\eta(t_i,x_{\mathrm{fit}})}{2\sigma_i^2}-\frac{1}{2}\log \frac{1}{\sigma_i^2}\Big).
\end{equation}
Assumptions were made that the resulting likelihood estimators for the three parameters are efficient, and that the number of data points are sufficient to be in the large sample limit, simplifying the estimation of uncertainties on the fit parameters, $x_{\mathrm{fit}}$.

%Under these conditions, the inverse of the covariance matrix may be estimated by computing the second derivative of the log-likelihood function. The inverse of the resulting matrix can then be taken to get the uncertainty on each of the parameters. 

The likelihood fitting process was used to fit each model to a combined data set of \acrshort*{sdom}1, 3 and 5, which were shown individually in Fig. \ref{fig:master_plot}. This data set was combined by taking the average and variance of each month of measurements. Each data set was re-sampled 10000 times in order to estimate a likelihood distribution for that model. P-values for the two models were calculated from this and are shown in table \ref{tab:mc_bio}.

%A resampling based on each model was performed in order to and calculate a p-value for that model. Analysis of the individual \acrshort*{sdom}s showed stochastic fluctuations that were accounted for in the sampling by taking an average of the deviations on each month of data. 

Both of the models considered in the final analysis can be approximated by a linear fit about the inflection point. This line was extrapolated to the point where it intersects with a line indicating no change in efficiency. The time associated with this point is used to calculate a parameter that will be referred to as the critical time. 
\begin{equation}
    T_\textrm{c}= -\frac{1-(\eta(T_\textrm{i})-T\eta'(T_\textrm{i}))}{\eta'(T_\textrm{i})}.
\end{equation}
We chose to define the critical time this way in order to 
estimate where the optical data transitioned from steady to a decline, while accounting for the fact that different models were applied. The efficiency of each model was evaluated at the critical time 
and found to be around 90$\%$ for both the Gompertz and Logistic models.

%On future instruments, monitoring when the transparency of optical surfaces fall below 90$\%$ of their value at immersion can be used as an indicator of entering the rapid-growth phase.  

\begin{table*}
    \centering
    \caption{Table of parameters from a Logistic and Gompertz fit. The final level $A$, fouling rate  $k_{\textrm{foul}}$, inflection time $T_\textrm{i}$ and calculated critical time $T_\textrm{c}$ of the two models after fitting . Time measurements are taken from the start of data taking, 8 months after the initial deployment.}
   \begin{tabular}{c|c|c|c|c|c}
         Model & $A$  &$k_{\textrm{foul}}$ ($\%$ yr$^{-1}$) & $T_\textrm{i}$ (yr) & $T_\textrm{c}$ (yr)  &p-value of fit\\
       \hline
        Logistic &  0.77$\pm$0.12 &26.0$\pm$3.8 & 3.2$\pm$0.3&  1.72$\pm$0.15 & 0.92\\
        Gompertz &  1.14$\pm$0.37 &23.4$\pm$6.9 & 3.3$\pm$0.5&  1.54$\pm$0.21 &0.98\\
        \hline
    \end{tabular}
    \label{tab:mc_bio}
\end{table*}

\begin{figure}
    \centering
    \includegraphics[scale=0.35]{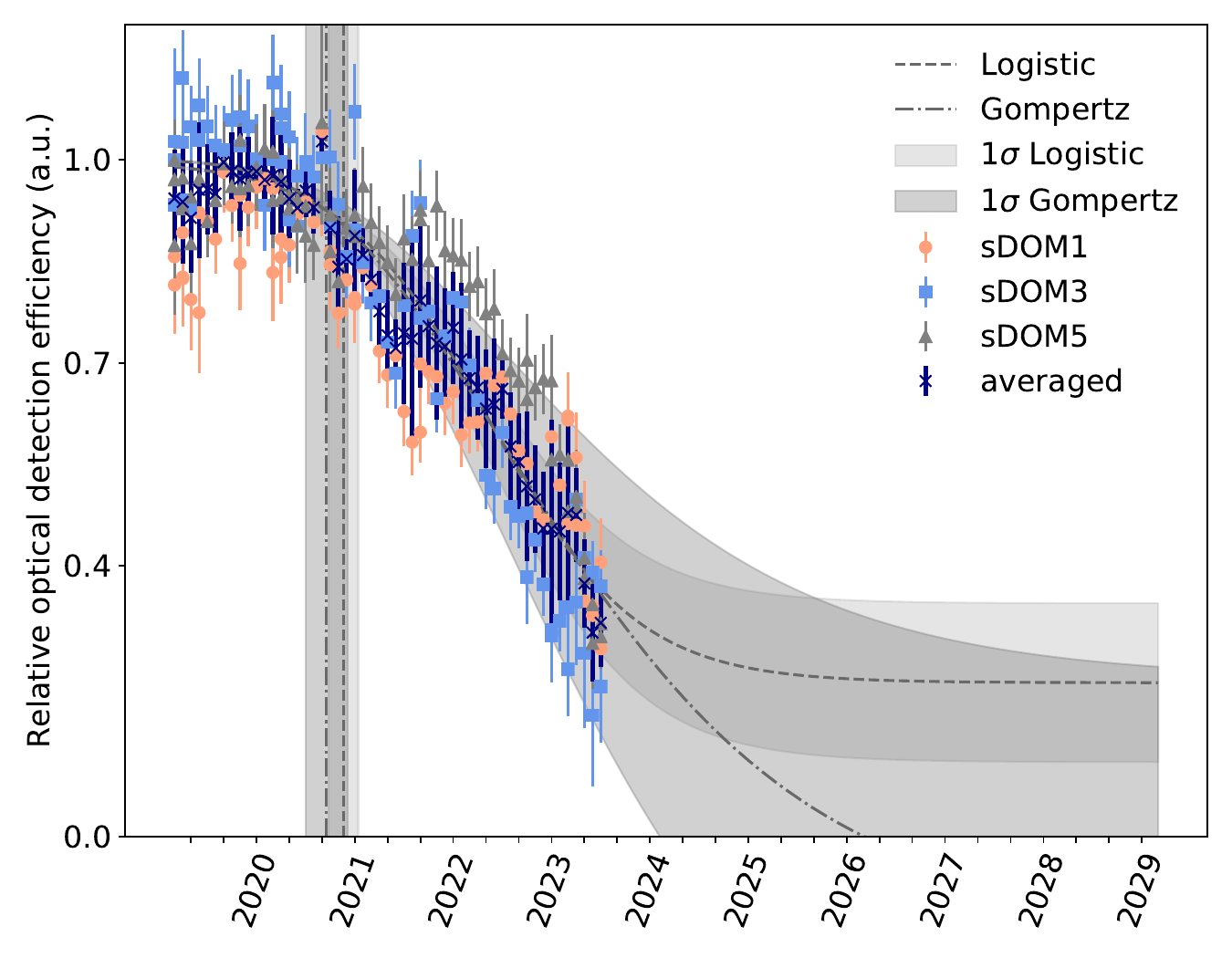}
    \caption{Fit of the logistic and Gompertz models to the combined data from Fig. \ref{fig:master_plot}. Rates of fouling are shown for each model and are in good agreement with each other. The vertical dashed and dashed-dotted lines represent the calculated critical times, and the shaded bands around these lines indicate the corresponding 1$\sigma$ confidence interval. Another confidence interval is shown around the expected maximal transparency loss. The fit is extrapolated in order to show the asymptotic behaviour of the fits and because future \acrshort*{pone} strings are being designed for a minimal operation time of 10 years.}
    \label{fig: fits}
\end{figure}

The critical time and fouling rate were calculated for the combined data set using each fitted model. Information from the fits is summarized in table \ref{tab:mc_bio}. Each model showed an acceptable goodness of fit based on the calculated p-values. The critical times are shown with errors in the shaded horizontal bands on Fig. \ref{fig: fits}, which suggests that the growth rates predicted by each model are compatible. One data point in Fig. \ref{fig: fits} shows efficiency $>$ 1.0. corresponding to August 2020. Looking at this month in Fig. \ref{fig:monthly_correlation}, the amount of bioluminescence light seen was greater in the upward-facing \acrshort*{pmts}, which explains this data point. Error-bars for the August 2020 measurement in Fig. \ref{fig:monthly_correlation} suggest it is within expected statistical fluctuations. No flasher data from this month is available to enable investigation of other possible effects, such as the cleaning effect described in section 2.3. 

These growth rates suggest a maximum annual loss in transparency of around 25$\%$ per year shortly after the critical time has been reached. \acrshort*{straw} had been deployed for about 8 months when regular data taking started, so this result implies that rapid efficiency losses take hold at around the two and a half year mark.
Using the upper 1$\sigma$ limit on $A$ in Table \ref{tab:mc_bio} and using Fig. \ref{fig: fits}, we can extrapolate the asymptotic behaviour of the biofouling. The final transparency of the upward-facing glass ranges from complete obscuration to 35$\%$ of the initial value in the long term limit. 
%Carsten statistics bits are bit out of place, replace with Likelihood model used, that is probably what people meant. 
%Refer back to table where we calculate the p-value and re-word. 

\section{Sampling of Biofouling and Sedimentation}

During the recovery of the two pathfinder instruments, samples were taken of the biofouling and sediment that had accumulated. These samples were taken from various points on each instrument. In-water sampling was done using a suction collection system connected to a holding area (suction sampler) on the \acrshort*{rov}. Additional samples were taken onboard the deck of the recovery vessel. 

The placement of instrumentation on the \acrshort*{straw} pathfinder was shown in Fig. \ref{fig:straw} \cite{Boehmer2019,Bailly2021}. Suction samples were taken from the tops of \acrshort*{sdom}1 and \acrshort*{sdom}5, located at 70 m above the seafloor. No samples were taken of the bottom of these modules, as they appeared to be free of biofouling. Samples were taken both on the top and bottom of \acrshort*{sdom}2, where biofouling was present. 

\acrshort*{strawb} was a 450 m tall mooring line, which was equipped with several different types of modules for making environmental measurements. The purposes and general layout of these instruments can be found in Fig. 2. of Ref. \cite{Holzapfel2024}. Suction samples were taken from the tops of the modules positioned at 432 m, 144 m, and 120 m above the sea floor.

\begin{figure*}
    \centering
    \includegraphics[width=1.0\linewidth]{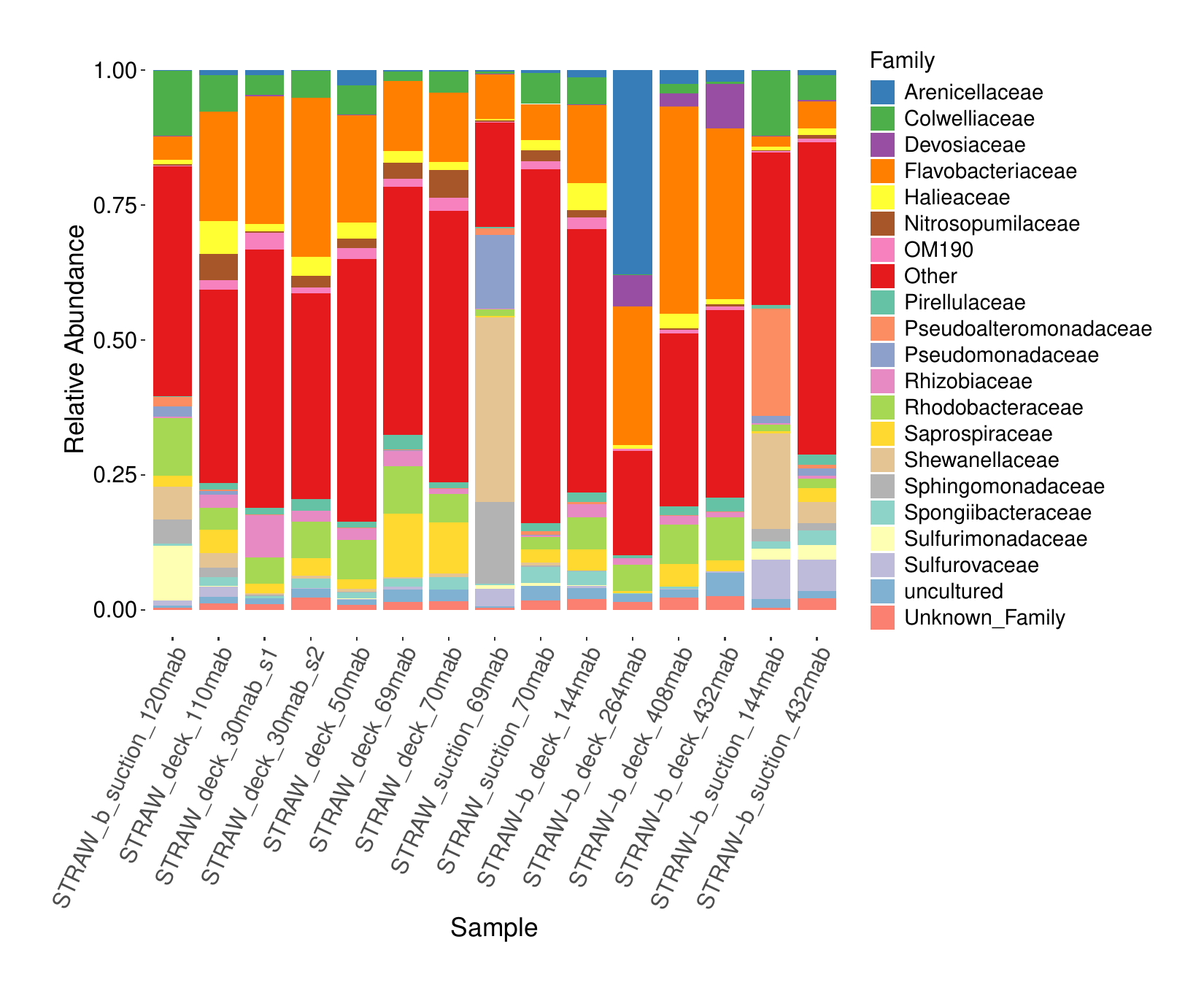}
    \caption{Stacked bar plot showing the relative abundance of the 20 most abundant families of each successfully amplified sample collected for microbiome analysis from \acrshort*{straw} and \acrshort*{strawb}. Sample names are labelled by the pathfinder apparatus (\acrshort*{straw} or \acrshort*{strawb}), sampling technique (suction or deck), and the meters above the sea floor (e.g, 144mab). Two measurements (s1 and s2) are shown for the STRAW instrument at 30mab, as it was sampled on both the top and bottom surfaces.
}
    \label{fig: microbes_samples}
\end{figure*}

The objective of taking samples was to identify the microbial component, known as the biofilm, of the accumulated biofouling. Identifying these could lead to mitigation strategies for biofouling, which will improve the long term light collection efficiency of upward-facing modules. Figure \ref{fig: microbes_samples} shows the microbial families identified on sampled modules.

DNA was extracted from the biofilm samples using the Zymobiomics DNA miniprep kit (Zymo Research, Irvine, CA). Polymerase chain reaction (PCR) amplification was carried out according to the protocols of the \textsf{Earth Microbiome Project} (see Ref. \cite{EMP}) following Ref. \cite{Thurber2020}. Amplicons were sequenced on an Illumina MiSeq (V.2 chemistry and 2x250 paired-end sequencing) by Oregon State University’s Center for Quantitative Life Sciences (CQLS). Sequences were quality-filtered by \textsf{FASTP} using default settings \cite{Chen2023}. 16S analysis was carried out on forward reads using the \textsf{Qiime2 Dada2} wrapper to predict Amplicon Sequence Variants (ASVs) \cite{Estakietal202}, using the \textsf{Silva v138.1 n99 database}. Samples were then rarefied to the minimum read count of the successfully amplified samples (51659 reads), and filtered based on at least 4 reads of a taxa present in 10$\%$ of samples. Visualization and analysis were carried out using \textsf{MicrobiomeAnalyst} \cite{Chongetal2020}, and the \textsf{phyloseq} R package \cite{McMurdieandHolmes2013}.

\begin{figure*}
    \centering
    \includegraphics[width=1.0\linewidth]{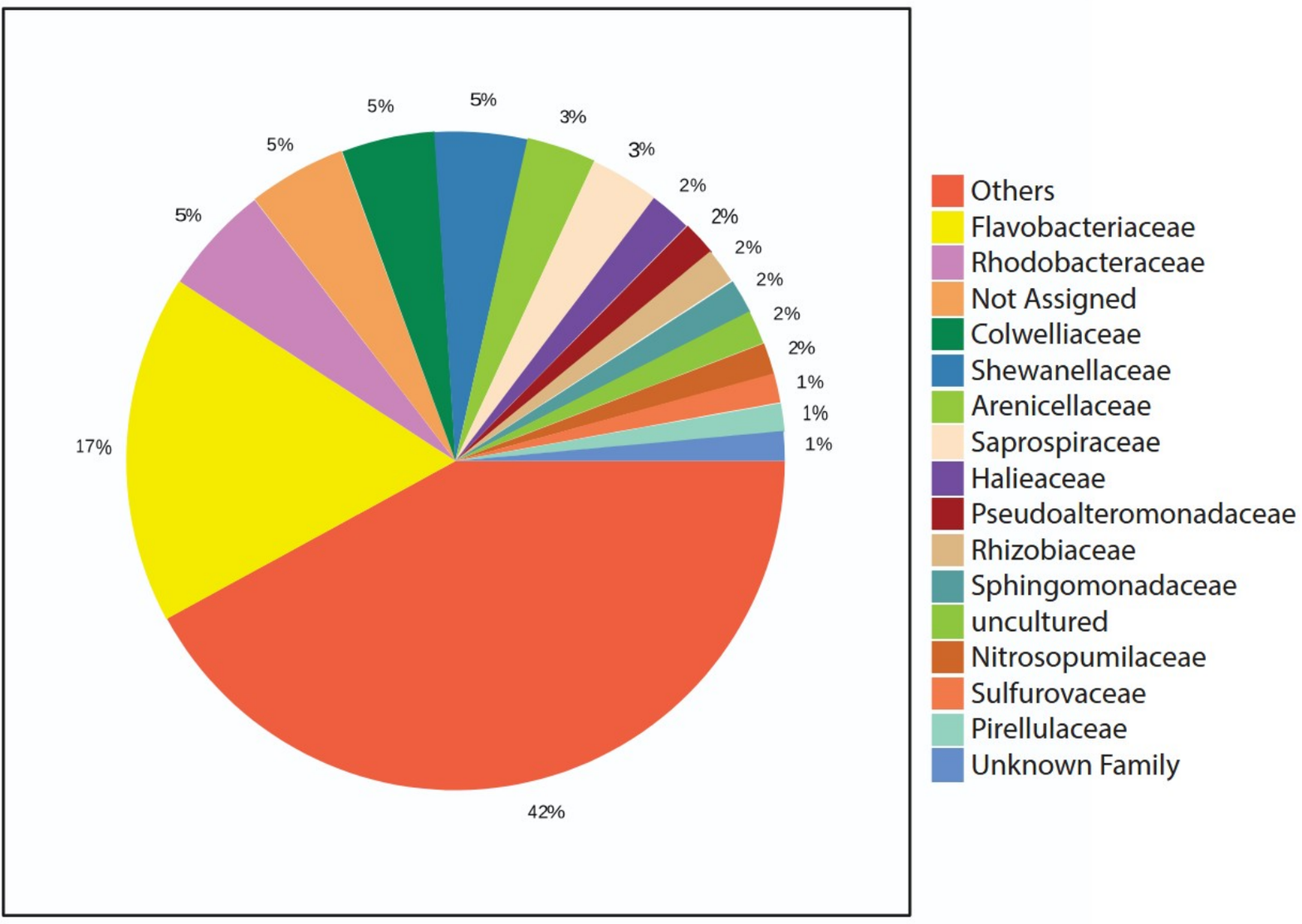}
    \caption{Pie chart showing the relative abundance of the 15 most abundant bacterial and archaeal families across all successfully amplified samples. Families outside of the 15 most abundant are grouped into the ‘Others’ category.  Samples were taken from both \acrshort*{straw} and \acrshort*{strawb} modules across varying depths and sensor types. 
}
    \label{fig: families_piechart}
\end{figure*}

When examining the microbial diversity of the microbial biofilms, we found them to be highly heterogeneous (Fig. \ref{fig: microbes_samples}) and dominated by heterotrophic microbes (Fig. \ref{fig: families_piechart}). The most abundant taxa were members of the family Flavobacteriaceae (17$\%$), rhodobacteriaceae (5$\%$), colwelliaceae (5$\%$),  and shewanellaceae (5$\%$) (Fig. \ref{fig: families_piechart}); all of these families are largely heterotrophic, using organic matter from the environment to obtain their energy \cite{Bowman2014,ChenandDing2023,Xiaetal2021}. As the dominant form of food at these depths is sinking organic matter that accumulates on surface-facing surfaces, angling sensors and developing methods where particulates do not collect would likely reduce biofilm formation on the sensors.

%All \acrshort*{pmts} on the \acrshort*{pone} optical modules are angled away from the vertical, so we expect the impact of biofouling to be less than what we have presented here for the purely upward-facing \acrshort*{straw} modules. Additionally, a subset of modules on the first P-ONE string will be deployed with ClearSignal$^\mathrm{TM}$ anti-biofouling coating, in order to test it's efficacy on upward-facing modules. 

\section{Discussion} \label{discussion}

\subsection{Results of This Study}

53 months of data from the P-ONE pathfinder was analyzed for indications that the transparency of the instrument's optical surfaces had changed. After around the two and a half year mark, the upward-facing glass rapidly became less transparent as shown in Fig. \ref{fig: fits}. This loss of efficiency is attributed to the effects of sedimentation and biofouling. The data available doesn't allow us to make a conclusion at what point this efficiency loss reaches equilibrium. Models motivated by relevant literature have been fit to the data and used to estimate the critical time and extrapolate a range for the final transmission efficiency. A scenario where the upward-facing light sensors become completely obscured cannot be ruled out based on the results of this fitting. A survey of the STRAW apparatus prior to recovery suggests that downward-facing optical modules are much less affected by fouling, and in several cases showed none. The sensitivity of the full \acrshort*{pone} detector to astrophysical sources will be dominated by up-going events, and therefore influenced primarily by partially downward-facing \acrshort*{pmts}. For example, Ref. \cite{NGC1068} used only up-going events in their analysis. 

The second pathfinder, \acrshort*{strawb} was surveyed by an \acrshort*{rov} before recovery, three years after deployment, suggesting that the impact of fouling decreases with distance above the seafloor. This is likely due to decreased re-suspension of organic and inorganic particulates, as well as some of the biofouling settling invertebrate larvae when moving away from the benthic boundary layer. However, during recovery of the \acrshort*{strawb} instrumentation fouling accumulation could not be quantified directly, so this is based on the qualitative observations made both from the \acrshort*{rov} surveying videos as well as in the laboratory aboard the recovery vessel. 

%The microbial biofilms on the \acrshort*{straw} and \acrshort*{strawb} optical modules aligns with biofilm community structure on the other researched deep-sea neutrino detector.

%The KM3NeT-ORCA site in the Mediterranean Sea has much reduced surface seasonal variability in comparison to the Cascadia Basin site studied here, although they share similar light attenuation characteristics at depth \cite{Bailly2021,Riccobene2007}. 

\acrshort*{straw} and \acrshort*{strawb} optical modules 
had similar gross microbial biofilm composition to the Mediterranean Sea (deep Ionian Sea) as reported in Refs \cite{Bellou2012,Bellou2011}. Both the deep Ionian and the STRAW biofilms included dominant groups of Gammaproteobacteria, Alphaproteobacteria, and Bacteroidia  (grouped into Flexibacter/ Cytophaga-Bacteroides in 2012). However, especially at shallower depths (yet still $>$1500m), there was a greater proportion of cyanobacteria in the Mediterranean Sea than we observed in the present study. Overall, this high level of similarity is remarkable considering the significant advance in microbial characterization since the important work of Ref. \cite{Bellou2012}. This supports a commonality in microbial biofilm communities within deep sea habitats even on arrays on different sides of the planet and vastly different oceanography. This basic understanding of community composition further allows a more directed study on the rate of microbial growth of the taxonomic groups that form the biofilm.  Growth rate studies based on this information can quantify a relationship between surface productivity, seasonality, and depth on the rate of biofilm formation since the general taxa appear, at least at a first order, to be similar across the oceans at these depths.

\subsection{Impact on P-ONE}

P-ONE-1, the first prototype line for the experiment, is currently under construction. 
This instrument will have 16 \acrshort*{pmt} optical modules and instrument 1 km of the water column, allowing for more robust measurements of the effects of biofouling and sedimentation. Based on observations as part of this analysis and the pathfinder recovery, some design choices may already offer some mitigation against these effects. All \acrshort*{pmts} on the \acrshort*{pone} optical modules are angled away from the vertical, so we expect the impact of biofouling to be less than what we have presented here for the purely upward-facing \acrshort*{straw} modules. For P-ONE-1 and future \acrshort*{pone} moorings, only the module closest to the seafloor on each string will be at a depth comparable to the \acrshort*{straw} modules discussed in this work.

In order to maximize the long-term sensitivity of upward-facing light sensors, fouling mitigation strategies are being explored. Currently, the most favorable anti-fouling approach is a fouling release coating. This type of coating works by weakening the adhesion between biofouling organisms and the infrastructure, allowing for the removal of these organisms by ocean currents or other mechanical means \cite{Finlay2013,Lejars2012,Hu2020}. The current industry standard is to use silicone based coatings which are commercially available and reduce biofouling population adhesion by up to 97$\%$ \cite{Lobe2015}.
A subset of modules on the first P-ONE string will be deployed with ClearSignal$^\mathrm{TM}$ anti-biofouling coating, in order to assess it's impact on the design of the full \acrshort*{pone} array \cite{clearsignal_website}. On future instruments, monitoring when the transparency of optical surfaces fall below 90$\%$ of their value at immersion can be used as an indicator of entering a rapid biofouling phase.

\begin{acknowledgements}
We thank Ocean Networks Canada for the very successful operation of the NEPTUNE observatory, as well as the support staff from our institutions without whom this experiment and P-ONE could not be operated efficiently.  We acknowledge the support of the Natural Sciences and Engineering Research Council of Canada (NSERC) and the Canadian Foundation for Innovation (CFI). This research was enabled in part by support provided by the BC and Prairies DRI and the Digital Research Alliance of Canada (alliancecan.ca). This research was undertaken thanks in part to funding from the Canada First Research Excellence Fund through the Arthur B. McDonald Canadian Astroparticle Physics Research Institute. P-ONE is supported by the Collaborative Research Centre 1258 (SFB1258) funded by the Deutsche Forschungsgemeinschaft (DFG), Germany. We acknowledge support by the National Science Foundation. This work was supported by the Science and Technology Facilities Council, part of the UK Research and Innovation, and by the UCL Cosmoparticle Initiative. This work was supported by the Polish National Science Centre (NCN).
\end{acknowledgements}

\bibliographystyle{spphys}       % APS-like style for physics
\bibliography{biofouling_paper}   % name your BibTeX data base

%\begin{thebibliography}{}
% etc
%\end{thebibliography}

\end{document}